\documentclass[oneside,10pt]{article}

\usepackage{graphicx}
\usepackage{changepage}

\newenvironment{myenv1}{\begin{adjustwidth}{-2.0cm}{-2.0cm}}{\end{adjustwidth}}

\usepackage{array}

\newcolumntype{x}[1]{%
 >{\centering\hspace{0pt}}p{#1}}%

\begin{document}
\title{Comparison of Sinogram-based Iterative Reconstruction with Compressed Sensing Techniques in X-ray CT}
\author{Dragos Trinca and Eduard Libin\\{\large Spitalul Clinic Judetean de Urgenta Arad, Romania}}
\date{}
\maketitle
\begin{abstract}
Performing X-ray computed tomography (CT) examinations with less radiation has recently received increasing interest: in medical imaging this means less (potentially harmful) radiation for the patient; in non-destructive testing of materials/objects such as testing jet engines, the redution of the number of projection angles (which for large objects is in general high) leads to a substantial decreasing of the experiment time. In the experiment, less radiation is usually achieved by either (1) reducing the radiation dose used at each projection angle or (2) using sparse view X-ray CT, which means significantly less projection angles are used during the examination. In this work, we study the performance of the recently proposed sinogram-based iterative reconstruction algorithm in sparse view X-ray CT and show that it provides, in some cases, reconstruction accuracy better than that obtained by some of the Total Variation regularization techniques. The provided accuracy is obtained with computation times comparable to other techniques. An important feature of the sinogram-based iterative reconstruction algorithm is that it has no parameters to be set.
\end{abstract}

\section{Introduction}
Iterative reconstruction algorithms \cite{BeisterM2012} have been extensively applied recently for the special case of sparse view X-ray computed tomography (CT) \cite{BeisterM2012}. Iterative algorithms have been used for a long time in tomography, but only during the last few years several manufacturers have made available and suggested the use of iterative techniques for medical CT imaging that simultaneously provide for good image quality with detectability of low-contrast lesions and significant dose reduction. In the special case of sparse view X-ray CT (that is, dose reduction by using significantly less projection angles, or views, during the examination) the iterative reconstruction algorithms are applied for obtaining a reconstruction of accuracy comparable with the case that uses an usual number of views.

Besides the optimization of the Algebraic Reconstruction Technique (ART), Simultaneous Algebraic Reconstruction Technique (SART), and Simultaneous Iterative Reconstruction Technique (SIRT) methods \cite{ART1,SART1,SIRT1}, the recent researches based on Total Variation (TV) regularization propose some of the most promising algorithms for sparse view X-ray CT \cite{ZhangH2016,TianZ2011,LiuY2012,CaiAL2014,ChangM2013,YangJS2010,NiuSZ2014,QiH2015,LiM2016}. However, all these papers report long computation times associated with TV regularization techniques. For example:
\begin{enumerate}
\item In \cite{ZhangH2016}, computation times are reported for reconstructions of size 256 by 256 pixels and 90 views; they are between 19 and 28 seconds.
\item In \cite{NiuSZ2014}, it is stated regarding the computation time, that "Similar to most SIR methods in CT image reconstruction (Ma et al, 2012a, Lauzier Thériault and Chen, 2013), the computational cost of the present PWLS-TGV method is very large because of the projection and back-projection operations using a huge system matrix. However, with the development of fast computers and dedicated hardware (Xu and Mueller, 2005, Knaup et al, 2006), iterative reconstruction algorithms, including the present PWLS-TGV method, may be used in clinical CT reconstruction in the near future." The authors did not report any computation time.
\item In \cite{QiH2015} the authors have proposed TV regularization techniques that are run with 100 iterations, and each iteration takes 44.33 seconds, for a reconstruction of size 512 by 512.
\end{enumerate}

Besides this, most papers developing TV regularization techniques report problems with low contrast structures that tend to be smoothed out by the TV regularization, posing a great challenge for these techniques.

In \cite{HanX2011}, which is representative of another set of works based on POCS (projection onto convex sets) for sparse view X-ray CT, the authors report a number of comparisons but do not report any computation time; instead it is said: "computation time for any reconstruction algorithm and, in particular, for iterative algorithms can be of concern. Accelerated computation can be achieved by streamlining/parallelizing the algorithms and/or by exploiting the available, or rapidly available, high-performance commodity computational hardware such as multi-core CPU and graphic processing units".

Other important works include: the works of Fessler et. al \cite{RamaniS2012,KimJK2012,KimD2013} that propose some methods for accelerating the iterative reconstructions with focus on statistical iterative reconstruction; the works of Bouman et. al. \cite{YuZ2011,SabneA2017} that propose fast methods for model-driven X-ray CT; and the works of Unser et. al that propose second-order extensions of TV regularization \cite{DoganZ2011}.

In \cite{TrincaD2016-1} the performance of the recently proposed sinogram-based iterative reconstruction algorithm was briefly studied on a Shepp-Logan phantom. In this work, we study in more detail the performance of proposed sinogram-based iterative reconstruction algorithm, with focus on the case of sparse view X-ray CT.

It is shown that the sinogram-based iterative reconstruction algorithm provides good reconstruction accuracy (in some cases obtaining accuracy slightly better than that obtained by some variants of the TV regularization technique), with computation times comparable with those reported for the TV regularization techniques. The examples studied are: the NCAT thoracic cross-section studied also in other recent works, and then an image showing a pipe welding process cross-section.
\section{Sinogram-based iterative reconstruction}
\subsection{Review}
First, we review the sinogram-based iterative reconstruction (SbIR) algorithm \cite{TrincaD2016-1}. Let \textnormal{$\mu_{1}$} be the density matrix (of size $n_{x}$ lines by $n_{y}$ columns) to be reconstructed from sinogram $y$, $n_{d}$ the number of detectors and $n_{v}$ the number of views. If $M$ is $n_{d}*n_{v}$ and $N$ is $n_{x}*n_{y}$, then the system matrix $A$ is of size $M$ lines by $N$ columns and the sinogram $y$ is a vector of size $M$. If we write $\mu_{1}$ in vector format then $\mu_{1}$ is a vector of size $N$; the reconstruction problem is to find $\mu_{1}$ such that
\begin{displaymath}
A * \mu_{1} = y.
\end{displaymath}

There are two steps in the reconstruction process of the SbIR algorithm:
\begin{description}
\item[initialization:]
first we initialize the reconstruction $\mu_{1}$ with an approximation derived from the sinogram $y$ as follows:
\begin{enumerate}
\item[1.1] $\mu_{1}$ $\leftarrow$ $A'$ $*$ ($y$ $./$ $\alpha$)
\item[1.2] $\mu_{1}$ $\leftarrow$ $\mu_{1}$ $./$ $\beta$
\end{enumerate}
\item[iterations:] at each iteration, we update the reconstruction $\mu_{1}$ based on the difference between the sinogram $y$ and the current situation in $\mu_{1}$:
\begin{enumerate}
\item[2.1] $\tilde{y}$ $\leftarrow$ $A$ $*$ $\mu_{1}$
\item[2.2] $\mu_{2}$ $\leftarrow$ diag($\mu_{1}$) $*$ $A'$ $*$ ($y$ $./$ $\tilde{y}$)
\item[2.3] $\mu_{1}$ $\leftarrow$ $\mu_{2}$ $./$ $\beta$
\end{enumerate}
\end{description}
where:
\begin{itemize}
\item $A'$ is the transpose of $A$,
\item $./$ is the element-by-element division,
\item $\tilde{y}$ is the situation in the reconstruction $\mu_{1}$ at the current iteration, and is a vector of size $M$ as the sinogram $y$,
\item $\alpha=\sum_{j}A[i,j]$ is of size $M$,
\item $\beta=\sum_{i}A[i,j]$ is of size $N$;
\item each element of $\mu_{1}$ corresponds to a square of size 1.0 by 1.0 in the experiment;
\item we also consider that $y$ has been obtained as area integrals, at a Detector,View pair meaning that each of the elements that are traversed is traversed for an area $\leq$ 1.0; we could consider line integrals as well, but using area integrals seem to help in the understanding of the algorithm.
\end{itemize}

The algorithm is very natural and its workings can be explained as follows; first let's explain the initialization (when looking at the steps 1.1, 1.2 above, look only at the first element of $\mu_{1}$ which is $\mu_{1}[1]$):
\begin{itemize}
\item in step 1.1, each of the $M$ elements of $y$ $./$ $\alpha$ is the average sinogram measurement per unit of area traversed by the X-ray beam at the respective Detector,View pair; thus each element of $y$ $./$ $\alpha$ can be viewed as an approximation of each of the elements of $\mu_{1}$ that are traversed by the X-ray beam at the respective Detector,View pair;
\item so, in step 1.1 the first element of $A'$ $*$ ($y$ $./$ $\alpha$) is the sum of all approximations of $\mu_{1}[1]$ (from all Detector,View pairs in which $\mu_{1}[1]$ is traversed);
\item in step 1.2 $\mu_{1}[1]$ is divided by $\beta[1]$, where $\beta[1]$ is the sum of all non-zero elements in the first line of $A'$; this is natural, because each of the approximations of $\mu_{1}[1]$ that come from step 1.1 are for unit of area 1.0, but for some Detector,View pairs in which $\mu_{1}[1]$ is traversed it might be traversed only for some portion of its total area 1.0.
\end{itemize}

Given this explanation of the initialization step, a better writing of the algorithm would be:
\begin{description}
\item[initialization:]
first we initialize the reconstruction $\mu_{1}$ with an approximation derived from the sinogram $y$ as follows:
\begin{enumerate}
\item[1.1] $\mu_{1}$ $\leftarrow$ $B$ $*$ ($y$ $./$ $\alpha$)
\end{enumerate}
\item[iterations:] at each iteration, we update the reconstruction $\mu_{1}$ based on the difference between the sinogram $y$ and the current situation in $\mu_{1}$:
\begin{enumerate}
\item[2.1] $\tilde{y}$ $\leftarrow$ $A$ $*$ $\mu_{1}$
\item[2.2] $\mu_{1}$ $\leftarrow$ diag($\mu_{1}$) $*$ $B$ $*$ ($y$ $./$ $\tilde{y}$)
\end{enumerate}
\end{description}
where:
\begin{itemize}
\item $B$ is a matrix of size $N$ lines by $M$ columns, whose $i$-th line is obtained from the $i$-th line of $A'$ by dividing each element of the $i$-th line of $A'$ to $\beta[i]$.
\end{itemize}

Given this re-writing of the algorithm, the workings can be explained as follows (again look only at $\mu_{1}[1]$):
\begin{itemize}
\item in step 1.1, $\mu_{1}[1]$ is obtained (as a weighted sum) by multiplying the first line of $B$ with the vector of approximations $y$ $./$ $\alpha$; precisely, consider that there are $k$ ($k\leq{M}$) non-zero elements in the first line of $B$, let these be $B[1,j_{1}]$, $B[1,j_{2}]$,..., $B[1,j_{k}]$ where $1\leq{j_{1}}\leq{j_{2}}\leq...\leq{j_{k}}\leq{M}$; each of $B[1,j_{1}]$, $B[1,j_{2}]$,..., $B[1,j_{k}]$ corresponds to one of those Detector,View pairs that traverse $\mu_{1}[1]$, and represents the fraction (weight) from the total sum traversed $\beta[1]$; so we have
\begin{displaymath}
B[1,j_{1}] + B[1,j_{2}] + ... +  B[1,j_{k}] = 1.0;
\end{displaymath}
\item thus if in step 1.1 above $Y_{\alpha}$ is the vector of approximations $y$ $./$ $\alpha$, then we have
\begin{displaymath}
\mu_{1}[1] \leftarrow B[1,j_{1}] * Y_{\alpha}[j_{1}] + B[1,j_{2}] * Y_{\alpha}[j_{2}] + ... +  B[1,j_{k}] * Y_{\alpha}[j_{k}],
\end{displaymath}
which means that $\mu_{1}[1]$ is obtained as a weighted sum of approximations;
\item the same explanation goes for the steps 2.1 and 2.2 from each iteration: in 2.1, $\tilde{y}$ is the situation of the current iteration, and thus the vector $y$ $./$ $\tilde{y}$ is the vector of correction coefficients;
\item to see how $\mu_{1}[1]$ is computed in 2.2, observe that it is computed by multiplying the first line of the matrix diag($\mu_{1}$) $*$ $B$ with the vector of corrections $y$ $./$ $\tilde{y}$; thus if $Y_{\tilde{y}}$ is the vector of corrections $y$ $./$ $\tilde{y}$, then
\begin{displaymath}
\mu_{1}[1] \leftarrow B[1,j_{1}] * \mu_{1}[1] * Y_{\tilde{y}}[j_{1}] + B[1,j_{2}] * \mu_{1}[1] * Y_{\tilde{y}}[j_{2}] + ... +  B[1,j_{k}] * \mu_{1}[1] * Y_{\tilde{y}}[j_{k}],
\end{displaymath}
which means that at each iteration $\mu_{1}[1]$ is obtained as a weighted sum of approximations (the weights being again $B[1,j_{1}]$, $B[1,j_{2}]$,..., $B[1,j_{k}]$).
\end{itemize}
\subsection{A simple example}
Let $n_{x}=n_{y}=n_{d}=n_{v}=2$ (so $N=M=4$) and
\begin{displaymath}
\mu_{1}=\left[
\begin{array}{cc}
1.0 & 3.0\\
2.0 & 4.0
\end{array}
\right]
\end{displaymath}
the matrix to be reconstructed, written in vector format as
\begin{displaymath}
\mu_{1}=\left[
\begin{array}{c}
1.0\\
2.0\\
3.0\\
4.0
\end{array}
\right].
\end{displaymath}

\begin{figure}
\begin{myenv1}
\centering
\includegraphics{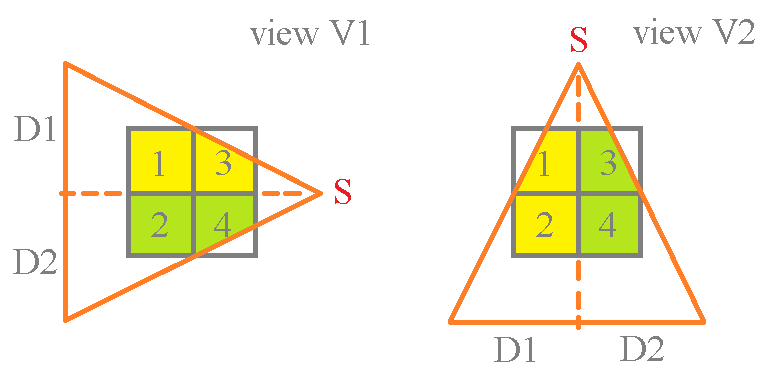}
\caption{\label{fig:experiment}at the left is the geometry of view V1, and at the right the geometry of view V2; in yellow is the area of $\mu_{1}$ that is traversed by that portion of the beam corresponding to detector D1 and in green is the area of $\mu_{1}$ that is traversed by that portion of the beam corresponding to detector D2}
\end{myenv1}
\end{figure}

We consider that the sinogram $y$ is obtained without any noise using the views V1 and V2, the view V1 at 0 degrees and the view V2 at a rotation around the object of 90 degrees, as in Figure \ref{fig:experiment}; if each of the 4 squares of $\mu_{1}$ is of size 1.0 by 1.0, and the distance from X-ray source S to center of rotation (i.e. the center of reconstruction $\mu_{1}$) is 2.0, and the distance from center of rotation to the detectors' line is also 2.0, and the detectors' line is of length 4.0, this means that:
\begin{itemize}
\item in view V1, $\mu_{1}[1]$ and $\mu_{1}[2]$ are each traversed by a full area of 1.0, whereas $\mu_{1}[3]$ and $\mu_{1}[4]$ are each traversed by an area of 0.75;
\item in view V2, $\mu_{1}[2]$ and $\mu_{1}[4]$ are each traversed by a full area of 1.0, whereas $\mu_{1}[1]$ and $\mu_{1}[3]$ are each traversed by an area of 0.75.
\end{itemize}

Thus
\begin{displaymath}
A=\left[
\begin{array}{cccc}
1.0 & 0.0 & 0.75 & 0.0\\
0.0 & 1.0 & 0.0 & 0.75\\
0.75 & 1.0 & 0.0 & 0.0\\
0.0 & 0.0 & 0.75 & 1.0
\end{array}
\right],
y=\left[
\begin{array}{c}
3.25\\
5.0\\
2.75\\
6.25
\end{array}
\right],
\end{displaymath}
\begin{displaymath}
\alpha=\left[
\begin{array}{c}
1.75\\
1.75\\
1.75\\
1.75
\end{array}
\right],
\beta=\left[
\begin{array}{c}
1.75\\
2.0\\
1.5\\
1.75
\end{array}
\right],
\end{displaymath}
\begin{displaymath}
B=\left[
\begin{array}{cccc}
\frac{1.0}{1.75} & 0.0 & \frac{0.75}{1.75} & 0.0\\
0.0 & \frac{1.0}{2.0} & \frac{1.0}{2.0} & 0.0 \\
\frac{0.75}{1.5} & 0.0 & 0.0 & \frac{0.75}{1.5}\\
0.0 & \frac{0.75}{1.75} & 0.0 & \frac{1.0}{1.75}
\end{array}
\right];
\end{displaymath}
the first line of $A$ corresponds to D1,V1, the second line to D2,V1, the third line to D1,V2, and the fourth line to D2,V2; also note that the fact that all values in $\alpha$ are equal is particular of this example (in other geometries they could be all different).

The SbIR algorithm works as follows (look only at $\mu_{1}[1]$):
\begin{itemize}
\item in 1.1, $\mu_{1}[1]$ is initialized by multiplying the first line of $B$ with the vector $y$ $./$ $\alpha$, so
\begin{displaymath}
\mu_{1}[1] = \frac{1.0}{1.75}*\frac{3.25}{1.75} + \frac{0.75}{1.75}*\frac{2.75}{1.75} = 1.734;
\end{displaymath}
note that $\frac{3.25}{1.75}$ is the average sinogram measurement per unit of area in the D1,V1 pair and $\frac{2.75}{1.75}$ is the average sinogram measurement per unit of area in the D1,V2 pair; also note that $\mu_{1}[1]$ is traversed in all Detector,View pairs for a total area of 1.75 (this is $\beta[1]$), with an area of 1.0 traversed in the D1,V1 pair and an area of 0.75 in the D1,V2 pair;
\item in 2.1, in the first iteration, the situation is
\begin{displaymath}
\tilde{y}=\left[
\begin{array}{c}
3.769\\
4.662\\
3.514\\
5.3
\end{array}
\right].
\end{displaymath}
\item in 2.2, in the first iteration, $\mu_{1}[1]$ is obtained by multiplying the first line of diag($\mu_{1}$) $*$ $B$ with the vector $Y_{\tilde{y}}$:
\begin{displaymath}
\mu_{1}[1] = \frac{1.0}{1.75}*1.734*\frac{3.25}{3.769} + \frac{0.75}{1.75}*1.734*\frac{2.75}{3.514} = 1.434;
\end{displaymath}
before this first iteration the situation in $\mu_{1}$ corresponding to the D1,V1 pair is 3.769, this means the elements that are traversed in D1,V1 need to be corrected by multiplication with $\frac{3.25}{3.769}$, so this is where the first approximation $1.734*\frac{3.25}{3.769}$ of $\mu_{1}[1]$ comes from; before this first iteration the situation in $\mu_{1}$ corresponding to the D1,V2 pair is 3.514, this means the elements that are traversed in D1,V2 need to be corrected by multiplication with $\frac{2.75}{3.514}$, so this is where the second approximation $1.734*\frac{2.75}{3.514}$ of $\mu_{1}[1]$ comes from;
\item as we can observe, $\mu_{1}[1]$ converges to its real value 1.0;
\end{itemize}
\section{Performance of the SbIR algorithm, with focus on sparse view X-ray CT}
We examine the performance of the SbIR algorithm by two examples, using the following scanning geometry for a fan-beam X-ray source, with detectors aligned in line:
\begin{enumerate}
\item the number of detectors is $n_{d}=1024$, the size of each detector being 1.41, so the detectors line has length 1448.15;
\item the distance from the X-ray source to the center of rotation is 1024.0, and the distance from the center of rotation to the detectors' line is also 1024.0;
\item the reconstruction is of size 512 by 512 pixels, with each pixel corresponding to a physical square area of size 1.0 by 1.0;
\item the number of views is $n_{v}=120$ covering the full circle, so the angle (in radians) between two consecutive views is $\frac{2\pi}{120}$.
\end{enumerate}

The SbIR algorithm is compared against the TV regularization techniques GPBB, UPN and UPN0 implemented in the TVReg software package developed recently at the Technical University of Denmark (DTU):
\begin{enumerate}
\item the Barzilai-Borwein accelerated gradient projection first-order method (GPBB);
\item implementation of an optimal first-order method for strongly convex functions, due to Nesterov, tailored to large-scale total variation regularization (UPN);
\item a first-order method by Nesterov, Beck, and Teboulle (UPN0) for the case of a zero strong convexity parameter.
\end{enumerate}

The TVReg software package is written in C language, and is available at the Github software repository \cite{TVReg-github} and described in \cite{TVReg-paper}. The TVReg package solves the problem
\begin{displaymath}
\textnormal{min}_{x} \{\frac{1}{2}||Ax - y||_{2}^{2} + \alpha\textnormal{TV}(x)\},
\end{displaymath}
where $x$ represents the reconstructed image, $y$ the sinogram data, and $A$ the forward operator (the system matrix). $\textnormal{TV}(x)$ is the Total Variation of the image, and $\alpha$ is a regularization parameter set by the user. Other parameter (besides $\alpha$) that can be set by the user is the noise parameter $rnl$ that is used to add Gaussian noise to the synthetically constructed sinogram $y$ in the case of simulated experiments.

Regarding the SbIR algorithm, the synthetically simulated sinograms in the two examples that we show are calculated using line integrals, as in most works in sparse view X-ray CT.
\subsection{Example 1}
\begin{figure}
\begin{myenv1}
\centering
\includegraphics[width=1.2\textwidth]{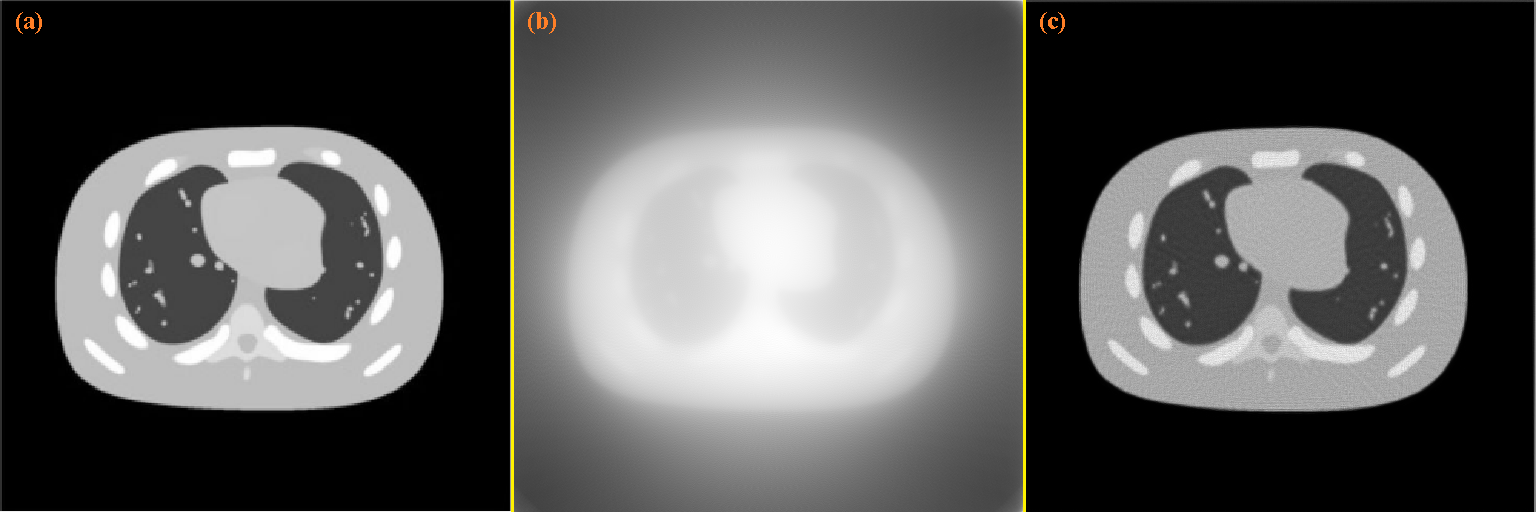}
\caption{\label{fig:1-originalPlusSbIR}(a) the original NCAT phantom, (b) the initialization of the SbIR algorithm and (c) the reconstruction with the SbIR algorithm after 512 iterations; both (b) and (c) are obtained after adding noise (with $rnl=0.005$) to the sinogram}
\end{myenv1}
\end{figure}

In the first example we consider the NCAT phantom studied also in other works \cite{TianZ2011} and shown in Fig. \ref{fig:1-originalPlusSbIR} (a) (this phantom simulates a thoracic cross-section); in this example we add noise to the sinogram by considering $rnl=0.005$ for both SbIR and the GPBB, UPN, UPN0 algorithms. Also the number of iterations is 512 for all algorithms (SbIR, GPBB, UPN, UPN0).

\begin{figure}
\begin{myenv1}
\centering
\includegraphics[width=1.2\textwidth]{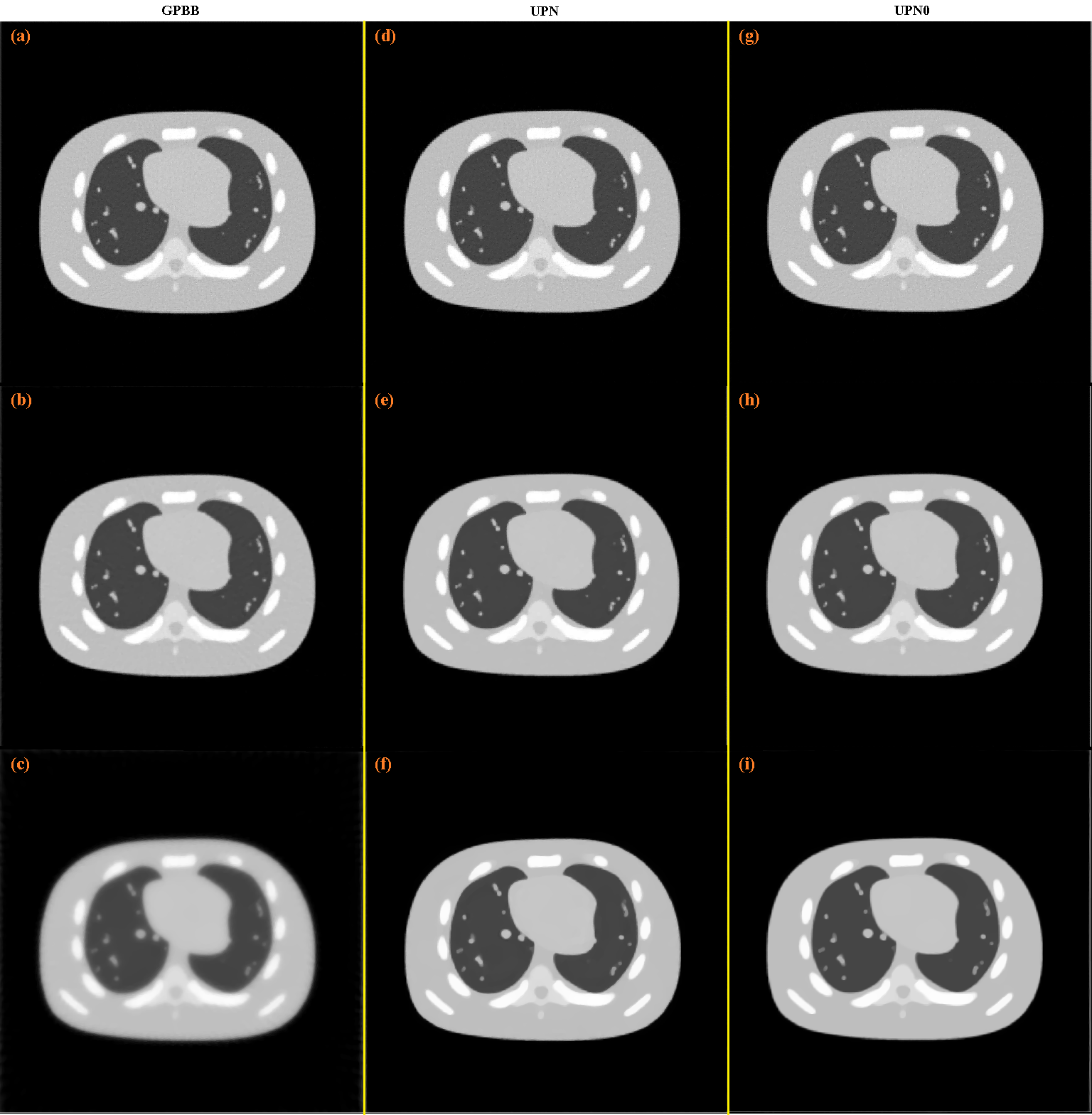}
\caption{\label{fig:TV-1}(a), (b) and (c) are the results after 512 iterations for the GPBB method for $\alpha=0.5$, $\alpha=5.0$ and $\alpha=50.0$ respectively; (d), (e) and (f) are the results after 512 iterations for the UPN method for $\alpha=0.5$, $\alpha=5.0$ and $\alpha=50.0$ respectively; (g), (h) and (i) are the results after 512 iterations for the UPN0 method for $\alpha=0.5$, $\alpha=5.0$ and $\alpha=50.0$ respectively}
\end{myenv1}
\end{figure}

The reconstruction with the SbIR algorithm is shown in Fig. \ref{fig:1-originalPlusSbIR} (c), where in Fig. \ref{fig:1-originalPlusSbIR} (b) is shown the reconstruction obtained after the initialization step is applied. The reconstructions with the GPBB, UPN, UPN0 algorithms are shown in Fig. \ref{fig:TV-1} when the regularization parameter $\alpha$ is 0.5, 5.0, or 50.0. In this example, the best reconstructions obtained with the GPBB, UPN, UPN0 algorithms are obtained when $\alpha=5.0$ (this means the middle row (b), (e) and (h) in Fig. \ref{fig:TV-1}).

We calculated the SSIM (Structural Similarity Index), PSNR (Peak Signal-to-Noise Ratio), MSE (Mean Square Error) for all the reconstructions: high values of SSIM and PSNR means higher accuracy while high values of MSE means lower accuracy; for the SbIR reconstruction they are SSIM:0.9018, PSNR: 25.6149, MSE: 0.0027; for GPBB, UPN, UPN0 they are listed in Table \ref{table:1-SSIM-PSNR-IMMSE}; as we can see the only case when the reconstruction with the SbIR algorithm is better is the one shown in Fig. \ref{fig:TV-1} (c).
\begin{table}
\caption{\label{table:1-SSIM-PSNR-IMMSE}Quantitative calculations showing the difference between each of the reconstructions with GPBB, UPN, UPN0 and the original, for example 1}
\centering
\begin{tabular}{|c|x{0.25\textwidth}|x{0.25\textwidth}|x{0.25\textwidth}|}
\hline
& GPBB & UPN & UPN0\tabularnewline\hline
$\alpha=0.5$ & SSIM: 0.9427, PSNR: 37.1377, MSE: 1.9330e-04 & SSIM: 0.9444, PSNR: 38.4690, MSE: 1.4226e-04 & SSIM: 0.9447, PSNR: 38.4914, MSE: 1.4154e-04 \tabularnewline\hline
$\alpha=5.0$ & SSIM: 0.9791, PSNR: 35.1269, MSE: 3.0712e-04 & SSIM: 0.9920, PSNR: 41.8835, MSE: 6.4812e-05 & SSIM: 0.9919, PSNR: 41.9283, MSE: 6.4146e-05 \tabularnewline\hline
$\alpha=50.0$ & SSIM: 0.8869, PSNR: 27.3581, MSE: 0.0018 & SSIM: 0.9690, PSNR: 32.1766, MSE: 6.0582e-04 & SSIM: 0.9725, PSNR: 32.6921, MSE: 5.3801e-04 \tabularnewline
\hline
\end{tabular}
\end{table}
\subsection{Example 2}
A problem with the GPBB, UPN, UPN0 algorithms is that for a low value of $\alpha$ the reconstruction is of low quality, whereas for a high value of $\alpha$ the reconstruction tends to become "foggy" and with visible patches; in example 1, the value $\alpha=5.0$ seems to be a good choice.

\begin{figure}
\centering
\includegraphics[width=0.4\textwidth]{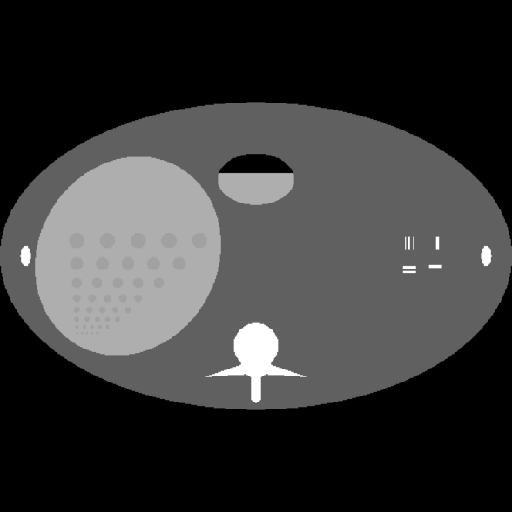}
\caption{\label{fig:2}abdominal phantom proposed by the Institute of Medical Physics, Friedrich-Alexander-University Erlangen-Nurnberg, Erlangen}
\end{figure}

We applied the same experimental setup but this time with $rnl=0.001$ (so with lower level of noise) to the abdominal phantom shown in Fig. \ref{fig:2}, proposed by the Institute of Medical Physics, Friedrich-Alexander-University Erlangen-Nurnberg, Erlangen \cite{pd1}; the reconstruction with UPN for this phantom was also best for $\alpha=5.0$ as in the case of example 1, whereas GPBB and UPN0 were best for $\alpha=0.5$ in this case. The conclusion would be that $\alpha=5.0$ works best in general for the UPN algorithm, for reconstructions of size 512 by 512 pixels; for GPBB and UPN0, the choice of $\alpha$ is more unclear.

\begin{figure}
\begin{myenv1}
\centering
\includegraphics[width=1.2\textwidth]{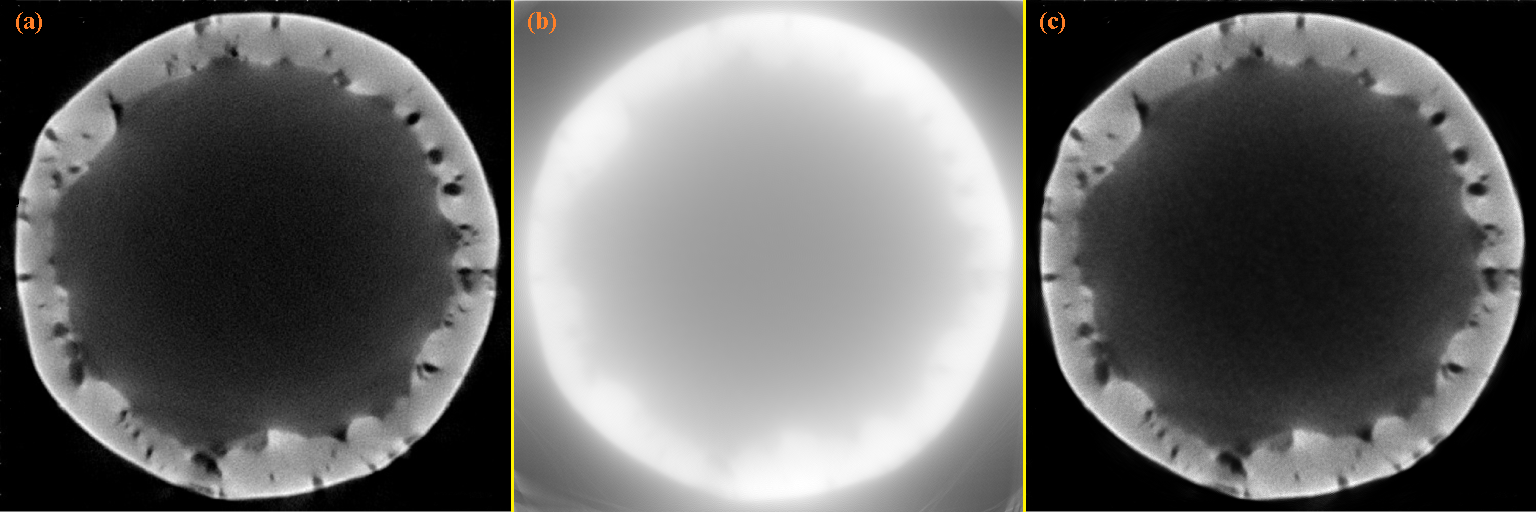}
\caption{\label{fig:4-originalPlusSbIR}(a) the original pipe welding process image, (b) the initialization of the SbIR algorithm and (c) the reconstruction with the SbIR algorithm after 512 iterations; both (b) and (c) are obtained after adding noise (with $rnl=0.001$) to the sinogram}
\end{myenv1}
\end{figure}

In this second example we show that setting $\alpha=5.0$ could lead to reconstruction obtained by the UPN algorithm of lower quality than the one obtained with the SbIR algorithm. Consider the image shown in Fig. \ref{fig:4-originalPlusSbIR} (a), which corresponds to a cross-section from a pipe welding process, image acquired during 2015 at the Institute of Non-destructive Testing, Tomsk Polytechnic University, Tomsk (this is an image from a non-destructive material testing experiment instead of medical imaging, but it is equally important since reducing the number of views yields an overall reduction of the experiment time); in this example, we add noise to the sinogram by considering $rnl=0.001$ for both SbIR and the GPBB, UPN, UPN0 algorithms. Also the number of iterations is 512 for all algorithms (SbIR, GPBB, UPN, UPN0).

\begin{figure}
\begin{myenv1}
\centering
\includegraphics[width=1.2\textwidth]{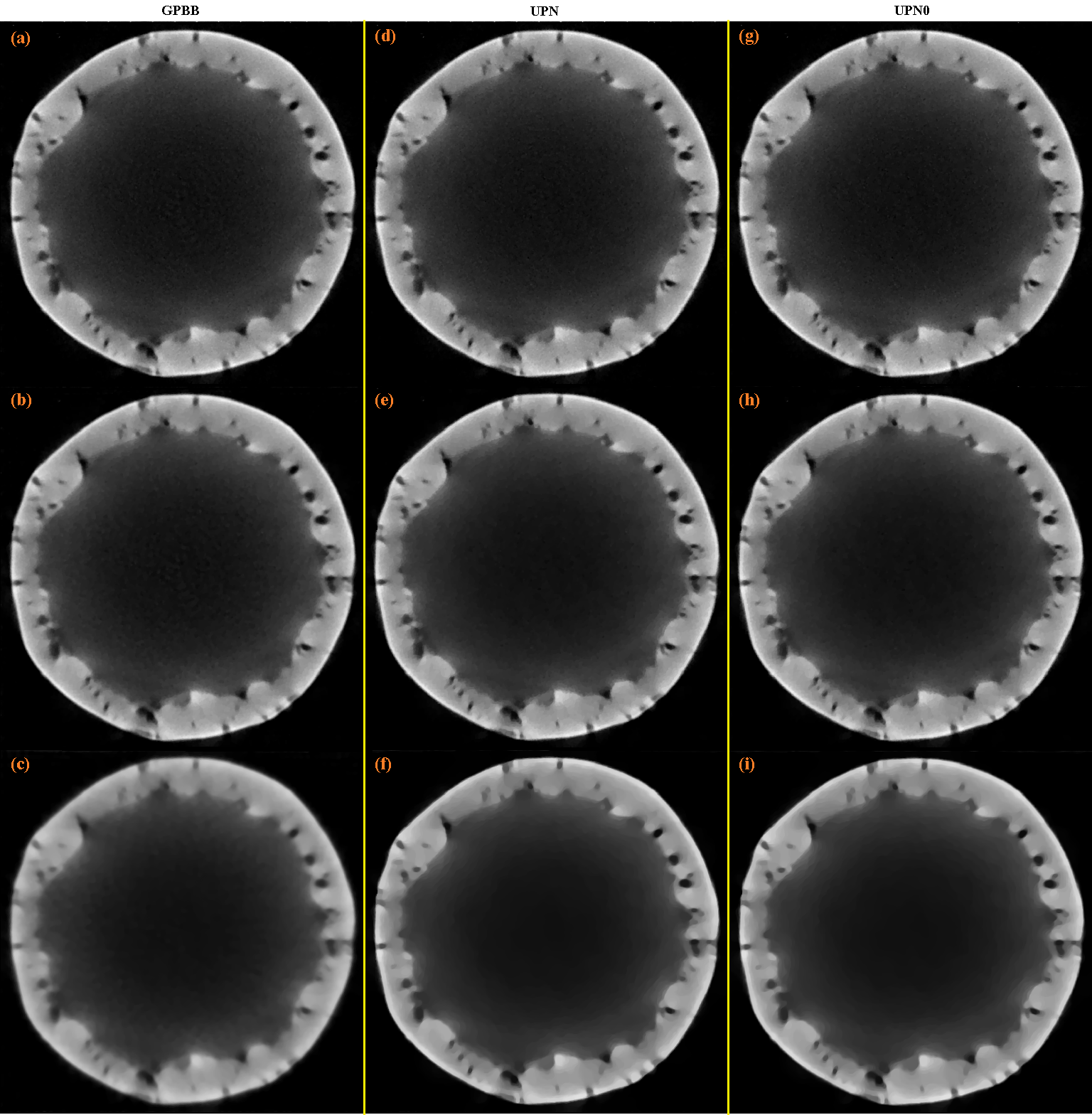}
\caption{\label{fig:TV-4}(a), (b) and (c) are the results after 512 iterations for the GPBB method for $\alpha=0.5$, $\alpha=5.0$ and $\alpha=50.0$ respectively; (d), (e) and (f) are the results after 512 iterations for the UPN method for $\alpha=0.5$, $\alpha=5.0$ and $\alpha=50.0$ respectively; (g), (h) and (i) are the results after 512 iterations for the UPN0 method for $\alpha=0.5$, $\alpha=5.0$ and $\alpha=50.0$ respectively}
\end{myenv1}
\end{figure}

The reconstruction with the SbIR algorithm is shown in Fig. \ref{fig:4-originalPlusSbIR} (c), where in Fig. \ref{fig:4-originalPlusSbIR} (b) is shown the reconstruction obtained after the initialization step is applied. The reconstructions with the GPBB, UPN, UPN0 algorithms are shown in Fig. \ref{fig:TV-4} when the regularization parameter $\alpha$ is 0.5, 5.0, or 50.0. The best reconstructions obtained with the GPBB, UPN, UPN0 algorithms are obtained when $\alpha=0.5$ (this means the first row (a), (d) and (g) in Fig. \ref{fig:TV-4}).

We calculated the SSIM, PSNR, MSE for all the reconstructions; for the SbIR reconstruction they are SSIM: 0.8820, PSNR: 35.5374, MSE: 2.7942e-04; for GPBB, UPN, UPN0 they are listed in Table \ref{table:4-SSIM-PSNR-IMMSE}; as we can see the reconstruction with the SbIR algorithm is better than all 6 reconstructions from Fig. \ref{fig:TV-4} (b), (c), (e), (f), (h), (i), but worse than the reconstructions corresponding to $\alpha=0.5$ shown in Fig. \ref{fig:TV-4} (a), (d), (g). Thus, this second example shows that even if in general the TV regularization techniques converge faster than the SbIR algorithm, there are cases as the one shown here where the SbIR algorithm provides a better reconstruction. Also, the SbIR algorithm has no parameters to be set as in the case of the TV regularization techniques.
\begin{table}
\caption{\label{table:4-SSIM-PSNR-IMMSE}Quantitative calculations showing the difference between each of the reconstructions with GPBB, UPN, UPN0 and the original, for example 2}
\centering
\begin{tabular}{|c|x{0.25\textwidth}|x{0.25\textwidth}|x{0.25\textwidth}|}
\hline
& GPBB & UPN & UPN0\tabularnewline\hline
$\alpha=0.5$ & SSIM: 0.9077, PSNR: 37.4524, MSE: 1.7979e-04 & SSIM: 0.9075, PSNR: 37.5485, MSE: 1.7585e-04 & SSIM: 0.9066, PSNR: 37.5471, MSE: 1.7591e-04 \tabularnewline\hline
$\alpha=5.0$ & SSIM: 0.8760, PSNR: 34.5804, MSE: 3.4831e-04 & SSIM: 0.8790, PSNR: 35.1570, MSE: 3.0500e-04 & SSIM: 0.8790, PSNR: 35.1281, MSE: 3.0703e-04\tabularnewline
\hline
$\alpha=50.0$ & SSIM: 0.7947, PSNR: 28.4132, MSE: 0.0014 & SSIM: 0.8372, PSNR: 31.7535, MSE: 6.6780e-04 & SSIM: 0.8382, PSNR: 31.8891, MSE: 6.4728e-04\tabularnewline
\hline
\end{tabular}
\end{table}
\subsection{Computation times achieved by the SbIR, GPBB, UPN, UPN0 algorithms}
The reconstructions have been obtained on a conventional PC with an i7-6500u @ 2.5 GHz 4-cores processor, and 8 GB of memory; only one core out of the four available has been used for each of the algorithms SbIR, GPBB, UPN, UPN0. The SbIR algorithm was implemented in C language in Visual C++ 2015; the GPBB, UPN, UPN0 algorithms of the TVReg package are implemented in C language. For the reconstructions shown in the two examples (all with 512 iterations) the computation time was around 150 seconds for the SbIR algorithm and between around 100 and 250 seconds for the GPBB, UPN, UPN0 algorithms. In \cite{TrincaD2016-2}, the complete code of our implementation in Visual C++ 2015 (using Win32 application, without any MFC or ATL code) of the SbIR 
algorithm can be found. In the code:
\begin{enumerate}
\item ID\_X\_ITERATIVEMETHOD1 is the id of the 'Iterative Method 1' menu option that runs the SbIR algorithm, with the geometry defined by the global variables.
\end{enumerate}
\subsection{Convergence of the SbIR algorithm}
\begin{figure}
\centering
\includegraphics{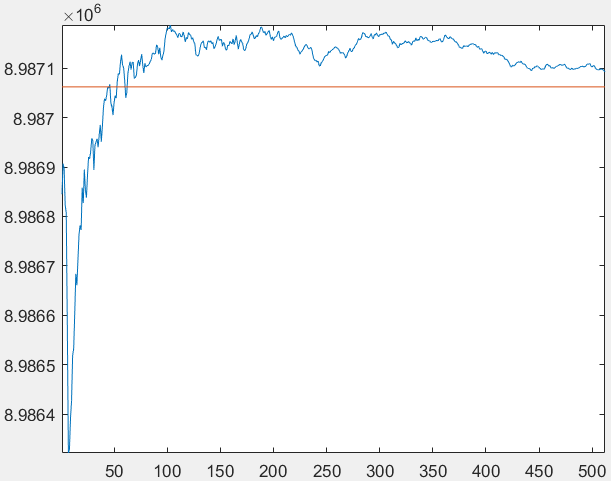}
\caption{\label{fig:4-convergence}convergence of the SbIR algorithm, example 2 (the pipe welding process image)}
\end{figure}

The convergence of the SbIR algorithm, after a number of studies we have done, seems to behave as follows: first the SbIR algorithm spends a number (that depends on the specific input and noise in the sinogram) of iterations to stabilize within a certain interval, after which the rest of the iterations are spent to converge to the optimal solution. For the second example (the pipe welding process image) that is run with 512 iterations, the convergence behavior is shown in Fig. \ref{fig:4-convergence}: the red graph shows the sum
\begin{displaymath}
\sum_{i=1}^{M}y[i]
\end{displaymath}
of all sinogram values, which for the second example it is $8.987062\times10^{6}$; and in blue graph it is plotted the sum
\begin{displaymath}
\sum_{i=1}^{M}\tilde{y}[i]
\end{displaymath}
of all situation values, at each of the 512 iterations. We observe that in this plot the SbIR algorithm spends around 100 iterations to stabilize within a certain interval, and after that starts to converge to an approximate solution.

We have tried other initial solutions for the SbIR algorithm (like the solution given by the Filtered Back-Projection algorithm) to see if with other initializations it converges faster, but it seems that its initialization as shown in this work works best.
\section{Conclusions}
From the results reported, the SbIR algorithm clearly proves to be a good option for obtaining good reconstruction accuracy in the special case of sparse view X-ray CT, in some cases with accuracy slightly better than some variants of the TV regularization technique. Comparing to the TV regularization techniques the SbIR algorithm has no parameters. One important work that needs to be done is proving the convergence of the SbIR algorithm mathematically. Other important further works include: (1) trying to find better initial solutions in order to speed up its convergence; and (2) development of optimal parallelizations using either multi-core processors, or Graphical Processing Units (GPUs), or both (hybrid CPU-GPU parallelizations) .

\end{document}